\documentclass{amsart}
\usepackage{amsmath}
\usepackage{diagbox}
\usepackage{stfloats}
\usepackage{enumerate}
\usepackage{float}
\usepackage{subfigure}
\usepackage{mathrsfs}
\usepackage{cases}
\usepackage{txfonts}
\usepackage{amsfonts}
\usepackage{mathrsfs}
\usepackage{amssymb}
\usepackage{graphicx}
\usepackage{tabularx}
\usepackage{caption}
\usepackage{siunitx}
\usepackage{multirow}
\usepackage{xmpmulti}
\usepackage{epstopdf}
\usepackage{booktabs}  
\usepackage{array}
\usepackage{colortbl,dcolumn}
\usepackage[utf8]{inputenc}

\newtheorem{theorem}{Theorem}[section]

\newtheorem{proposition}[theorem]{Proposition}
\newtheorem{corollary}[theorem]{Corollary}
\theoremstyle{definition}
\newtheorem{definition}[theorem]{Definition}

\theoremstyle{remark}
\newtheorem{remark}[theorem]{Remark}

\numberwithin{equation}{section}
\def\R{\mathbb{R}}

\begin{document}

\title[POMP]{Matrix Pre-orthogonal Matching Pursuit and Pseudo-Inverse}


\author{Wei Qu}
\address[Qu]{College of Sciences, China Jiliang University, China.}
\email{quwei2math@qq.com}

%
\author{Chi Tin Hon*}
\address[Hon]{Macau Institute of Systems Engineering, Macau University of Science and Technology, Macau, China}
\email{cthon@must.edu.mo}

\author{Yi Qiao Zhang}
\address[Zhang]{Macau Institute of Systems Engineering, Macau University of Science and Technology, Macau, China}
\email{zhangyqemily@163.com}

\author{Tao Qian*}
\address[Qian]{Macau center for mathematical sciences, Macau University of Science and Technology, China}
\email{tqian@must.edu.mo}

\thanks{*Corresponding author: Tao Qian and Chi Tin Hon}

\date{}

\dedicatory{}


\begin{abstract}

We introduce a new fundamental algorithm called Matrix-POAFD to solve the matrix least square problem. The method is based on the matching pursuit principle. The method directly extracts, among the given features as column vectors of the measurement matrix, in the order of their importance, the decisive features for the observing vector. With competitive computational efficiency to
 the existing sophisticated least square solutions
the proposed method, due to its explicit and iterative algorithm process, has the advantage of trading off minimum norms with tolerable error scales. The method inherits recently developed studies in functional space contexts. The second main contribution, also in the algorithm aspect, is to present a two-step iterative computation method for pseudo-inverse. We show that consecutively performing two least square solutions, of which one is to $X$ and the other to $X^*,$ results in the minimum norm least square solution. The two-step algorithm can also be combined into one solving a single least square problem but with respect to $XX^\ast.$ The result is extended to the functional formulation as well. To better explain the idea, as well as for the self-containing purpose, we give short surveys with proofs of key results on closely relevant subjects, including solutions with reproducing kernel Hilbert space setting, AFD type sparse representation in terms of matching pursuit, the general ${\mathcal H}$-$H_K$ formulation and pseudo-inverse of bounded linear operator in Hilbert spaces.
\end{abstract}

\maketitle

\tableofcontents
 \pagenumbering{arabic}

%

\section{Introduction}
We give a brief survey on the least square (LS)and Moore-Penrose generalized inverse (pseudo-inverse or PI) problems with the matrix and, generally, the Hilbert space settings. We first consider the matrix setting. Let us be given a pair of matrices $(X,Y),$ where $X=(x_{jk})$ is of order $m\times n$, called the {\it measurement matrix} or {\it kernel matrix}, and $Y$ of order $m\times p,$  called the {\it observing matrix}. The problem is to find an $n\times p$ matrix $W$ that solves the equation system
  \begin{eqnarray}\label{pseq}
  XW=Y.\end{eqnarray}
  This question is, in general, ill-posed, for there may not exist such $W$ to satisfy the exact identical relation. For this consideration we call $XW=Y$ as \emph{pseudo-equation}. There are two associated well-posed versions, of which one is known as the least square (LS) problem that is to find a matrix $W$ of the appropriate order such that the Frobenius norm of the difference matrix $XW-Y,$ denoted
\begin{eqnarray}\label{LS} \|XW-Y\|_2,\end{eqnarray}
is minimized. The second well-post version is to find an LS solution $W$ that is at the same time of the minimum norm among all the LS solutions of the pseudo-equation (\ref{pseq}). The problem can be solved through solving $p$ sub-questions, of which each is of the same type but for a column vector. From now on we assume that $p=1.$
\def\E{\mathcal E}
In this case, as is well known, from the classical least square theory, a solution $W_1$ to the LS problem is also a solution to the related \emph{normal equation}
\begin{eqnarray}\label{eq}X^*XW=X^*Y,\end{eqnarray} where $X^*$ is the transpose and complex conjugate of $X$ (\cite{BH}), and vice versa. We note that the normal equation (\ref{eq}) always has a solution, and usually has multiple solutions constituting a finite dimensional linear space.
In fact, basic linear algebra asserts that
 \begin{eqnarray}\label{assert} {\rm rank\ of\ matrix}\ X^\ast X= {\rm rank\ of\ the\ extended\ block\ matrix} \ (X^\ast X\ \ X^\ast Y),\end{eqnarray}
 which implies existence of solutions of the equation (\ref{eq}).

   Equivalence of the  minimization problem (\ref{LS}) and solution of equation (\ref{eq}) can be well explained by Hilbert space geometry. In fact, from (\ref{eq}) we have
 \begin{eqnarray}\label{ortho} X^\ast(XW-Y)=0.\end{eqnarray}
 This shows that if $W$ is a solution to (\ref{eq}), then the zero values of the product matrix exhibit that the column vector $XW-Y$ is orthogonal with all the column vectors of $X.$ This means that $XW-Y$ is perpendicular to the span of the column vectors of $X.$ Hence $XW,$  being a linear combination of the column vectors of $X,$ gives rise to the least distance.

  The equation (\ref{eq}) has a single solution  if and only if the columns vectors of $X,$
  denoted as $\vec{x}_k, k=1,\cdots,n,$ are linearly independent, and if and only if $X^*X,$ as an $n\times n$ matrix, has an inverse. In the case, the unique LS solution is given by $$W=(X^*X)^{-1}X^*Y.$$ It is, at the same time, of the minimum norm and thus the pseudo-inverse of $Y$ with respect to $X.$

 An LS problem has, and has only one solution under the constraint condition that the solution is of minimum norm among all the normal solutions of (\ref{eq}). If $W_1$ is a solution to (\ref{eq}) or equivalently to (\ref{LS}) and $W'$ is non-zero and $XW'=0,$ then $W_1+W'$ is also a solution to(\ref{eq}) or to (\ref{LS}) but with a norm larger than that of $W_1.$ In the general ${\mathcal H}$-$H_K$ formulation (see \S 4) it will be shown that $W_1$ is the minimum norm least square solution, or the pseudo-inverse if and only if (i) $W_1$ is a solution to the LS problem; and (ii) $W_1$ is in the orthogonal complement of the null space of the linear operator $L$ being defined through left multiplying by the matrix $X,$  i.e., $LW=XW, \forall W\in {\R}^n.$

Next we consider the functional space case in which similar questions concerning the LS and the pseudo-inverse may be asked. In the general cases, however, a solution may not exist. With the Hilbert space setting, a theory has been developed. There exists a sufficient and necessary condition under which the pseudo-inverse exists. The theory, in particular, asserts that existence of an LS solution implies existence of the pseudo-inverse (\cite{Na}). If, moreover, we work with a bounded linear operator in a reproducing kernel Hilbert space (RKHS), then the solution formula is explicitly presented {\cite{SS}. The matrix case well fits in the RKHS theory, for every finite dimensional Hilbert space is an RKHS.\\

We will first discuss the general Hilbert space case.
  Let $\mathcal{H}$ be a Hilbert space and $L$ a bounded linear operator from $\mathcal{H}$ to $H,$ where $H$ is a second Hilbert space containing the range $R(L)$ of $L$ as a subset. The least square (LS) problem for $Lf=d$ stands for the following:  For suitable $d\in H$ find $f\in \mathcal{H}$  that gives rise to
\begin{eqnarray}\label{RKHS} \inf_{f\in \mathcal{H}}\|Lf-d\|_{{H}}.\end{eqnarray}

Under the above condition, if moreover $(L^*L)^{-1}$ exists, then the unique and hence the minimum norm solution, denoted by $f_d,$ is given by
$$f_d=(L^*L)^{-1}L^*d.$$
 When a solution exists, $(L^*L)^{-1}$ may or may not exist.

If, further, ${\mathcal H}$ is a RKHS, i.e., ${\mathcal H}=H_K,$ with $K$ being its reproducing kernel, then without invertibility of $L^*L,$
the minimum norm LS solution, or the pseudo-inverse, can be explicitly expressed by
\begin{eqnarray}\label{direct} f_d(p)=\langle L^*d,L^*LK_p\rangle_{H_k},\end{eqnarray}
where $H_k$ is the induced RKHS identical in the set-theoretic sense with ${\rm ker}(L)^{\perp}=\overline{R(L^*L)}$. The new kernel $k$ is induced by the projection $P\ :\ H_K\to {\rm ker}(L)^{\perp},$ with the expression
\begin{eqnarray}\label{k}
k(p,q)=\langle L^\ast L K_q, L^\ast L K_p\rangle.
\end{eqnarray}

The operator $L^*$ can also be computationally realized. In fact, for any $d\in \mathcal{H}$ and $p\in \E,$
\[L^*d(p)=\langle L^*d,K_p\rangle_{H_K(E)}=\langle d,LK_p\rangle_{\mathcal{H}},\]
where the data $d,L$ and $K_p$ are all known (\cite{SS}).

The matrix case  well fits with the RKHS setting in which one
denotes by $L$ the bounded linear operator left-multiplying by the matrix $X.$
The matrix version of the condition $L^\ast d\in R(L^\ast L)$ is just (\ref{eq}). With the criterion (\ref{ad}) a matrix LS problem is always solvable. In fact, there hold the identical relations
\begin{eqnarray}\label{adj}L^*d\in {R(L^\ast)}=\overline{R(L^\ast)}=\overline{R(L^\ast L)}=R(L^* L),\end{eqnarray}
where we used Proposition \ref{prop} and finiteness of the dimensions.

In summary, each of the relations (\ref{eq}), (\ref{assert}), (\ref{ortho}), (\ref{ad}), (\ref{or}) is sufficient to ensure existence of a solution to the LS problem in the respective context (\ref{LS}) or (\ref{RKHS}).

We recall that pseudo-inverse is a further question to the LS problem: Seek for $\tilde{f}\in {\mathcal H}$ with minimum norm giving rise to
\begin{eqnarray}\label{noRKHS} \inf_{f\in {\mathcal H}}\|Lf-d\|_{{H}}.\end{eqnarray}

In \S 4 we will deal with the problem with the so called ${\mathcal H}$-$H_K$ formulation. It turns out that existence of an LS solution implies existence of the pseudo-inverse solution in all cases.

The contribution of the present study is 3-fold. One is to introduce Matrix-POAFD as a fundamental algorithm in solving the LS problem (\ref{LS}). It, specifically, and for the first time, extends the AFD methodology to the matrix context: The problem is raised and to be solved in the matrix form. In contrast, the preceding AFD methods are all for functional spaces. We note that although matching pursuit is employed at a technical point, the algorithm does not follow the general concept of greedy algorithm: It does not depend on the concept of dictionary. Without concerning the magnitudes of $m$ and $n,$ nor on ranks of the related matrix operators, the proposed method gives rise to a solution of (\ref{LS}). The second contribution is to solve the matrix form (\ref{noRKHS}), viz., the \emph{minimum norm least mean square problem}. In light of the general ${\mathcal H}$-$H_K$ theory we show that by consecutively solving two LS problems we can get the pseudo-inverse solution. Furthermore, the two-step algorithm can be combined into a one-step algorithm applied to the product matrix $XX^\ast.$
The third contribution is to generalize the two-step and one-step matrix pseudo-inverse algorithms to pseudo-inverse for operators in Hilbert spaces. In the general operator case an extra condition has to be imposed.  \\

The writing plan of the paper is as follows. \S 2 presents a revision on the two most known matching pursuit methods, viz., \emph{greedy algorithm} (GA) and \emph{orthogonal greedy algorithm} (OGA), as well as \emph{pre-orthogonal adaptive Fourier decomposition} (POAFD) that promotes Matrix-POAFD. We study approximation efficiency of the concerned algorithms that evidences optimality of POAFD in the sense of one-step energy matching.  In \S 3, we introduce Matrix-POAFD method together with algorithmic details. In \S 4, we review ${\mathcal H}$-$H_K$ formulation, and use it to solve pseudo-inverse problems in both the matrix and the general operator contexts. In \S 5, through numerical experiments we make comparison regarding to algorithmic efficiency between Matrix-POAFD and a number of mature methods (LSQR,CG,Ridge,MP).

\section{A Survey on Algorithms Based on Matching Pursuit Methodology}

Matrix-POAFD that we are to introduce belongs to a series of well studied algorithms in which the partial terminology AFD is an abbreviation of \emph{adaptive Fourier decomposition}. Greedy algorithms are with the general Hilbert space setting, while AFD is with the concrete complex analytic Hardy space setting, being a complex analysis method of harmonic analysis. In virtue of complex analysis, especially with properties of Blaschke products, AFD results in the Gabor type positive analytic instantaneous frequency decomposition. Like the greedy type algorithm, AFD, too, involves adaptive selections of parameters at its iterative steps. It, however, can always attain the global maximal energy matching pursuit at each of its iterative steps, and give rise to \emph{positive analytic instantaneous frequency decomposition} in the form of a fast converging Takenaka-Malmquist series (\cite{QWa}).  The type of expansions then be generalized, called \emph{pre-orthogonal adaptive Fourier decomposition} (POAFD), to any Hilbert space with a dictionary satisfying the so called \emph{boundary vanishing condition} (BVC). See (\ref{BVC}) or \cite{Q2D}. Further, the method was generalized to random signals, called \emph{stochastic AFD} (SAFD) and \emph{stochastic POAFD} (SPOAFD), taking into account of the stochastic properties of the signals under study (\cite{SAFD,WWRQW}).  The just mentioned S-type AFD variations are studied in the Bochner type spaces, being product spaces of a probability $L^2$-type and a Lebesgue $L^2$-type.  In both the theoretical and the practical studies of random fields, the stochastic AFD methods (S-type AFD) are competitive with the dominating Karhunen-Lo\`{e}ve expansion \cite{QZLQ} and PCA (principal component analysis) (see, for instance, \cite{DPO}).  AFD also has a stochastic $n$-best version with the classical Hardy space (\cite{QQ}), the latter being further extended to weighted Bergman and weighted Hardy spaces (\cite{QZLQ,Q2023}). Besides those separately cited, we also refer to \cite{Qianbook,QSW,CQT,DT,BCDD}) for further details, as well as applications.

   For a dense subset ${\mathcal L}$ of a Hilbert space ${\mathcal H}$  one can perform energy matching pursuit step by step to obtain linear combinations of elements of ${\mathcal L}$ to approximate a prescribed element of ${\mathcal H}.$ The set ${\mathcal L}$ is usually parameterized as
   \[ {\mathcal L}=\{e_q\ :\ q\in {\mathcal E}\},\] called a \emph{pre-dictionary}, and \emph{dictionary} if the elements of ${\mathcal L}$ are of norm $1.$

 Below we assume $\|e_q\|=1, \forall q\in {\mathcal E}.$ We will review three main types of matching pursuit algorithms: (i) general (or plain) greedy algorithm (GA, \cite{DT,Te}); (ii) orthogonal greedy algorithm (OGA), or equivalently, orthogonal matching pursuit (OMP), see \cite{DT,Te,PRK}); and (iii) pre-orthogonal adaptive Fourier decomposition (POAFD). POAFD is a generalization of AFD (\cite{QWa}) applicable to any Hilbert space with a dictionary satisfying the so called \emph{boundary vanishing condition} (BVC) (\cite{QSW,Q2D,Qu-Sps-represt-dirac,Qianbook}), that is
 \begin{eqnarray}\label{BVC} \lim_{q\to \partial {\mathcal E}}\langle f,e_q\rangle =0, \quad f\in {\mathcal H}.\end{eqnarray}
 In view of the Riemann-Lebesgue Lemma in classical Fourier analysis and the same property of Szeg\"o kernel in the original AFD, BVC is a natural condition to be imposed. When the parameter set  ${\mathcal E}$ is an open set, the optimal matching pursuits in the corresponding greedy algorithms at the iterative steps may not be reachable. Under the assumption of BVC, POAFD enjoys attainability of global optimality at each of its matching pursuit steps.

Below we sketch and compare the above mentioned three types of matching pursuit algorithms. We will show that, in terms of reconstruction efficiency, POAFD outperforms OGA, and OGA outperforms GA.

  Let $f\in {\mathcal H}$ be given and ${\mathcal L}$ a set of elements in ${\mathcal H}$ of unit norm.\\

(i) GA is based on an optimal ${\mathcal L}$-element selection with respect to the standard remainders:
\[ {q_n}=\arg\sup \{ |\langle r_{n},e_q\rangle|\ :\ q\in {\mathcal E}\},\]
where $r_{n}$ is the standard remainder formed iteratively from one by one optimal selections of $e_{q_1},\cdots,e_{q_{n-1}},$  if feasible.
 Precisely, $r_1=f,$
\[ r_2=f-\langle r_1,e_{q_1}\rangle e_{q_1}, \quad r_3=f-\langle r_1,e_{q_1}\rangle e_{q_1}-\langle r_2,e_{q_2}\rangle e_{q_2},\cdots\]

(ii) OGA (or OMP) is based on optimal ${\mathcal L}$-element selections, if feasible, with respect to the orthogonal remainders:
\begin{eqnarray}\label{last} {q_n}=\arg\sup \{ |\langle h_{n},e_q\rangle|\ :\ q\in {\mathcal E}\},\end{eqnarray}
where $h_{n}$ is the orthogonal remainder, being the difference between $f$ and the orthogonal projection of $f$ into the span of the already selected $e_{q_1},\cdots,e_{q_{n-1}}, h_1=f.$ Precisely, denote by $Q$ the  orthogonal complement of the corresponding projection which by itself is also a projection, we have
\[ h_2\triangleq f-\langle f,e_{q_1}\rangle e_{q_1}, \quad h_3\triangleq\left(I-{\rm Proj}_{{\rm span}\{e_{q_1},e_{q_2}\}}\right)(f)\triangleq Q_{{\rm span}\{e_{q_1},e_{q_2}\}}(f),\cdots\]

(iii) POAFD is based on optimal selections from the elements in the orthogonal complement in ${\mathcal L}$ of the span of the already selected with respect to, again, the orthogonal remainders $h_n,$
\[ {q_n}\triangleq\arg\max \left\{ \left|\left\langle h_{n},\frac{Q_{{\rm span}\{e_{q_1},\cdots,e_{q_{n-1}}\}}(e_q)}{\|Q_{{\rm span}\{e_{q_1},\cdots,e_{q_{n-1}}\}}(e_q)\|}\right\rangle\right|\ :\ q\in {\mathcal E}\right\}.\]
In POAFD \emph{boundary vanishing condition} (BVC) is assumed, and, as a consequence, in each matching pursuit step an optimal orthogonal complement ${\mathcal L}$-element is attainable.

The fact that OGA outperforms GA is based on the observation $\|h_3\|\le \|r_3\|,$ the latter being a consequence of the minimal distance given by the orthogonal projection in the Hilbert space. To compare POAFD with OGA, by using the properties $Q_{{\rm span}\{e_{q_1},\cdots,e_{q_{n-1}}\}}^2=Q_{{\rm span}\{e_{q_1},\cdots,e_{q_{n-1}}\}}, Q_{{\rm span}\{e_{q_1},\cdots,e_{q_{n-1}}\}}^\ast =Q_{{\rm span}\{e_{q_1},\cdots,e_{q_{n-1}}\}}$  and $h_n=Q_{{\rm span}\{e_{q_1},\cdots,e_{q_{n-1}}\}}(f),$ there hold
\[\left|\left\langle h_{n},\frac{Q_{{\rm span}\{e_{q_1},\cdots,e_{q_{n-1}}\}}(e_q)}{\|Q_{{\rm span}\{e_{q_1},\cdots,e_{q_{n-1}}\}}(e_q)\|}\right\rangle\right|=\frac{|\langle h_{n},e_q\rangle|}{\|Q_{{\rm span}\{e_{q_1},\cdots,e_{q_{n-1}}\}}(e_q)\|}\ge
|\langle h_{n},e_q\rangle|.\]
By comparing with (\ref{last}), the last inequality shows that the POAFD optimal selection gains more energy than OGA from the same orthogonal remainder $h_n$ (see also \cite{LQ}). Taking $n=3$ we observe that at the third step POAFD outperforms OGA.

 \section{Matrix-POAFD}

 The naming of Matrix-POAFD is to show the concept development. Besides its high efficiency compared with the existing methods in solving the LS problem, as a particular merit, it extracts out, in the order of their importance to the observing $Y,$ the decisive features represented by the columns of $X$. Since the algorithm is iterative, one can trade off the minimum norm scales with the tolerable errors.

With the matrix context the density issue of ${\mathcal L}$ reduces to one regarding the linear dependency.  As a result, a solution for the matrix question refers to whether there exists a shortest distance from $Y$ to the linear span of the column vectors of $X,$ and whether, as a solution of LS, the shortest distance is attainable, and whether the solution of minimum norm, as pseudo-inverse, is attainable.  The solution $W$ is usually not unique. What mostly happens is that $W_1\ne W_2$ but $XW_1=XW_2,$ giving rise to the same projection of $Y$ into ${\rm span}{\mathcal L}.$ In the matrix case, by the same reasoning as in the last section, the below described Matrix-POAFD outperforms other types of matching pursuit methods.

Let $\vec{u},\vec{v}$ be two column vectors of dimension $m,$ where $\|\vec{u}\|=1.$ Set
\[ Q_{\vec{u}}(\vec{v})=\vec{v}-\langle \vec{v},\vec{u}\rangle \vec{u}.\]
$Q_{\vec{u}}(\vec{v})$ is called \emph{the Gram-Schmidt ($G$-$S$) orthogonalization}, or \emph{orthogonal complement projection} (co-projection), of $\vec{v}$ onto $\vec{u}.$
We will also define the projection operator onto the span of a collection of vectors ${\mathcal A},$ denoted as ${\rm Proj}_{{\rm span}\{\mathcal A\}}(\vec{v}).$ In such form we have $Q_{\vec{u}}(\vec{v})={\rm Proj}_{{\rm span}\{\vec{u}\}}(\vec{v}).$

For any system of mutually orthogonal unit vectors $\{\vec{u}_1,\cdots,\vec{u}_k\},$
consecutively applying the GS orthogonalization, we have
\[ Q_{\vec{u}_k}(Q_{\vec{u}_{k-1}}(...(Q_{\vec{u}_1}(\vec{v}))\cdots )= Q_{\vec{u}_k}\circ Q_{\vec{u}_{k-1}}\circ \cdots Q_{\vec{u}_1}(\vec{v})=
\vec{v}-{\rm Proj}_{{\rm span}{\{\vec{u}_k,\cdots,\vec{u}_1\}}}(\vec{v}).\]

We now solve the LS problem (\ref{LS}) by using Matrix-POAFD. First we express $X$ by its column vectors,
\[ X=X^{(1)}=(\vec{x}^1_1,\cdots, \vec{x}^1_n).\]

Select a $\vec{x}^1_{k_1}$ that satisfies
\begin{eqnarray}\label{rule0} \left|\left\langle Y, \frac{\vec{x}^1_{k_1}}{\|\vec{x}^1_{k_1}\|}\right\rangle\right|
=\max\left\{\left|\left\langle Y, \frac{\vec{x}^1_k}{\|\vec{x}^1_k\|}\right\rangle\right|: k=1,\cdots,n, \vec{x}_k^{1}\ne 0\right\}.\end{eqnarray}
We note that there may be more than one column vectors $\vec{x}^1_{k}$ giving rise to the maximal projection.
Set $\vec{u}_1= \frac{\vec{x}^1_{k_1}}{\|\vec{x}^1_{k_1}\|}$ and construct an induced measurement matrix
\[ X^{(2)}=(\vec{x}^2_1,\cdots, \vec{x}^2_n),\]
where $\vec{x}^{2}_k=Q_{\vec{u}_1}(\vec{x}^1_k), \ k=1,\cdots,n.$ If  some $\vec{x}^1_k$ is parallel (proportional) to $\vec{u}_1,$ then the corresponding term $\vec{x}^2_k$ in $X^{(2)}$ is zero. In particular, $\vec{x}^2_{k_1}=0.$

Next, we select a vector $\vec{x}_{k_2}^{2}$ according to
\begin{eqnarray}\label{rule1} \left|\left\langle Y, \frac{\vec{x}^{2}_{k_2}}{\|\vec{x}^{2}_{k_2}\|}\right\rangle\right|
=\max\left\{\left|\left\langle Y, \frac{\vec{x}^{2}_k}{\|\vec{x}_k^{2}\|}\right\rangle\right|: k=1,\cdots,n; \vec{x}_k^{2}\ne 0 \right\}.\end{eqnarray}

Naturally, $k_2\ne k_1.$ If there is no $\vec{x}_k^{2}\ne 0$ to enable the selection,  then the process stops. In the case the best approximation to $Y$ by linear combinations of the column vectors of $X$ is just ${\rm Proj}_{{\rm span}\{\vec{u}_1\}}(Y),$ being a constant multiple of $\vec{x}^1_{k_1}.$ Otherwise, one can continue and set $\vec{u}_2= \frac{\vec{x}_{k_2}^2}{\|\vec{x}^{2}_{k_2}\|}$ and construct the next generation measurement matrix as
\[ X^{(3)}=(\vec{x}^3_1,\cdots, \vec{x}^3_n),\]
where
$\vec{x}^{3}_k=Q_{\vec{u}_2}(\vec{x}^{2}_k), k=1,\cdots,n.$ We note that $X^{(3)}$ contains at least two zero column vectors, namely  $\vec{x}^{3}_{k_1}, \vec{x}^{3}_{k_2}.$
We then select $\vec{x}^{3}_{k_3},$ if available, to give rise to the non-zero maximal correlation with $Y$ analogous to (\ref{rule0}) and (\ref{rule1}), and set $\vec{u}_3= \frac{\vec{x}_{k_3}^2}{\|\vec{x}^{2}_{k_3}\|},$ and so on.  This process will stop when getting, for the first time, an induced matrix $X^{(L+1)}, L+1\leq n,$ whose columns are either zero, or non-zero but orthogonal to $Y.$
 We note that in the iterative process we have kept the original order of the column vectors $\vec{x}_{k}$ in all the induced $X^{(l)}, l\leq L.$ \\

 In such a way, Matrix-POAFD results in an orthonormal system giving rise, in the order of importance, to all the contributive features of the observing vector $Y:$

\[ \vec{u}_l=\frac{Q_{\vec{u}_{l-1}}\circ \cdots \circ Q_{\vec{u}_1}(\vec{x}^{1}_{k_{l}})}{\|Q_{\vec{u}_{l-1}}\circ \cdots \circ Q_{\vec{u}_1}(\vec{}x^{1}_{k_{l}})\|}, \quad l=1,\cdots, L.\]\\

\begin{theorem}\label{POMP}
Matrix-POAFD algorithm results in an orthonormal system giving rise to an LS solution $W$ of (\ref{LS}).
\end{theorem}

\noindent{Proof}.
If $L$ reaches $n,$ then we have a normalized and mutually orthogonal system $\vec{u}_1,\cdots,\vec{u}_n$ such that
\[ {\rm Proj}_{{\rm span}\{\vec{x}^1_1,\cdots,\vec{x}^1_n\}}(Y)=\sum_{j=1}^{n}
\langle Y, \vec{u}_j\rangle \vec{u}_j,\]
and the LS of (\ref{LS}) is
\[ \|Y-XW\|^2=\|Y-UA\|^2=\|Y\|^2-\sum_{j=1}^{n}
|\langle Y, \vec{u}_j\rangle|^2\ge 0,\]
 where $U=(\vec{u}_1,\cdots,\vec{u}_n)$ and $A$ is the column vector $(\langle Y, \vec{u}_1\rangle,\cdots,\langle Y, \vec{u}_n\rangle)^\top.$ In the case $X$ is of rank $n.$ Letting $XW=UA,$
 we have the unique least square solution
$W=(X^\ast X)^{-1}X^\ast UA.$

In the case $L<n$ we selected an $L$-orthonormal system  $\vec{u}_1,\cdots,\vec{u}_{L},$ corresponding to the $k_1,\cdots,k_L$'th columns in the original matrix $X,$ and the rest columns are either zero or orthogonal to $Y.$ Set the index sets $K_1=\{k_1,\cdots,k_L\}$ and $K_2=\{1,\cdots,n\}\setminus K_1.$ We have
 \[ {\rm span}\{\vec{x}^1_{1},\cdots,\vec{x}^1_{n}\}={\rm span} \{\vec{u}_1,\cdots,\vec{u}_{L},\vec{x}_{i'_1},\cdots,\vec{x}_{i'_{n-L}}: i'_1,\cdots,i'_{n-L}\in K_2\}.\]
  The vectors $\vec{x}_{i'_1},\cdots,\vec{x}_{i'_{n-L}}$ correspond to the columns of $X$ that are in the orthogonal complement of the span of $\vec{u}_1,\cdots,\vec{u}_{L}$ and have no contributions to reconstruction of $Y.$ A solution $W=(w_1,\cdots,w_n)^T$ of the least square problem is separated into two parts corresponding to two steps: First we let $w_k=0$ if $k\in K_2.$ The rest $w_k$'s, viz. those for $k\in K_1,$ will be determined by the following process. Let $a_k=\langle Y,\vec{u}_k\rangle.$ Denote $U=(\vec{u}_1,\cdots,\vec{u}_{L}), A=(a_1,\cdots,a_L)^T,  \tilde{X}=(\vec{x}_{k_1},\cdots, \vec{x}_{k_L})$ and $\tilde{W}$ a column matrix of order-$L$ such that
 \[ \tilde{X}\tilde{W}=UA.\]
 It is then easy to see $\tilde{W}=(\tilde{X}^\ast \tilde{X})^{-1}\tilde{X}^\ast UA,$ as $\tilde{X},$ and hence $\tilde{X}^\ast \tilde{X}$ as well, have rank $L.$
 Denote $\tilde{W}=(\tilde{w}_1,\cdots,\tilde{w}_L)^T.$ Then the $w_k$'s for $k\in K_1$ are determined by setting $w_{k_j}=\tilde{w}_j, j=1,\cdots,L.$ Such defined $W$ is indeed a solution, because in writing $Y=Y_1+Y_2,$ where $Y_1$ is the projection of $Y$ into the span of the columns of $X,$ we have
 \begin{eqnarray}\label{ref} Y_1={\rm Proj}_{{\rm span}\{\vec{x}_1,\cdots,\vec{x}_n\}}(Y)=\sum_{j=1}^{L}
\langle Y, \vec{u}_j\rangle \vec{u}_j=UA=\tilde{X}\tilde{W}=XW,\quad Y_2\perp Y_1.\end{eqnarray}
This concludes that
$\|Y_2\|_2$ is the least square of (\ref{LS}), and
\[ \|Y-XW\|^2_2=\|Y_2\|_2^2=\|Y\|_2^2-\sum_{l=1}^L |a_l|^2.\]
 The proof is complete.\\

\begin{remark}
We note that, if the solution is not unique, then Matric-POAFD gives one of the multiple solutions. Hence, a such obtained solution $W$ may not be of minimum norm. This situation corresponds to when the columns of $X$ are linearly dependent. We give the following trivial example. Let $X=(\vec{x}_1, \vec{x}_2)\in {\R}^m\times {\R}^2,$ and $Y\in {\R}^2,$ where $\vec{x}_1, \vec{x}_2$ and $Y$ are all non-zero and parallel to each other. In this case there are multiple solutions $W=(w_1,w_2)$ such that $Y=XW.$ The pseudo-inverse solution with the setting (\ref{noRKHS}) will be specially addressed in the next section.
\end{remark}

\begin{remark}
A solution obtained through Matrix-POAFD may be regarded as a sparse approximation to $Y$ in using column vectors of $X$ which, due to the optimal column vector selection strategy, often gives fast convergence. In an ample amount of literature a different sparsity concept is used that refers to a large amount of zero coefficients. See, for instance, \cite{LW}, and references therein. In essence, however, the two concepts are similar: Both mean that not a full basis is used but only part.  Our concept of sparsity is not restricted to the matrix case, but extendable to functional spaces.
\end{remark}

%

\section{ ${\mathcal H}$-$H_K$ Formulation and Minimal Norm Solution}
We will first review the ${\mathcal H}$-$H_K$ formulation that, as a fundamental Hilbert space operator structure, is essentially contained in \cite{SS}. See also \cite{QHHK}.

Let
${\mathcal H}$ be a Hilbert space, say of the form ${\mathcal H}=L^2(D),$ that contains a family of kernel functions $h_p$ parameterized by $ p\in \E.$ The family $\{h_p\}_{p\in \E}$ induces an operator $L$ on ${\mathcal H}$ given by
\[ Lf(p)=\langle f,h_p\rangle_{\mathcal H}, \quad f\in {\mathcal H},\ p\in \E.\]
We alternatively write $F(p)=Lf(p).$
We assume the Hilbert-Schmidt boundedness condition $$M=\int_{\E}\|h_p\|_{{\mathcal H}}^2dp=\int_{\E}\int_{D}|h_p(x)|^2dxdp<\infty.$$

  Denote by
\[ N(L)=\{ f\in {\mathcal H}\ :\ Lf=0\}\]
 the null space of the operator $L.$
As a closed subspace of ${\mathcal H},$  $N(L)$  has an orthogonal complement in ${\mathcal H},$ denoted as $N(L)^\perp.$  There holds the direct sum decomposition ${\mathcal H}=N(L)^\perp \oplus N(L),$ being realized as $f=f^++f^-, f^+\in N(L)^\perp,
f^-\in N(L)$ and $Lf=Lf^+, \forall f\in {\mathcal H}.$ Recall that $R(L)$ denotes the range of operator $L$ defined on $H.$
There exists the following result involving the adjoint operator $L^\ast$.\\

\begin{proposition}\label{LSexist}
There exists a solution $f\in \mathcal{H}$ for (\ref{RKHS}) if and only if
\begin{eqnarray}\label{ad}
L^*d\in R(L^*L)\end{eqnarray}
(\cite{SS}), where $L^*$ is the adjoint operator of $L$ and $R(L^*L)$ stands for the range of the operator $L^*L.$ Moreover, in this case the pseudo-inverse solution exists.\end{proposition}
Proofs may be found in, for instance, \cite{SS} and \cite{Na}. For the self-containing purpose we include our short proofs.\\
\noindent{Proof}.
First, if $L^*d\in R(L^*L),$ then there exists $\tilde{f}\in {\mathcal H}$ such that $L^*d=L^*Lf.$ This implies
\begin{eqnarray}\label{or} L^\ast (d-L\tilde{f})=0,\end{eqnarray}
or, equivalently, $\forall g\in {\mathcal H},$
\[ 0=\langle g,L^\ast(d-L\tilde{f})\rangle=\langle Lg,d-L\tilde{f}\rangle,\ \  {\rm or}\ \ \ (d-L\tilde{f}) \perp R(L).\]
The orthogonality implies that $\tilde{f}$ gives rise to  $\|d-L\tilde{f}\|,$ being the LS. The above argument is reversible. In fact, if $\tilde{f}$ gives rise to the shortest distance  $\|d-L\tilde{f}\|,$ then $d-L\tilde{f}$ is orthogonal with $R(L).$ This final assertion means $L^\ast (d-L\tilde{f})=0,$ or $L^\ast d\in R(L^\ast L).$ Next we show that in such a case the minimum norm LS solution always exists. In fact if $\tilde{f}$ is an LS solution, then its orthogonal complement projection  $\tilde{f}^+ \in {\mathcal H}$ is also a solution and of the minimum norm. The proof is complete.\\

 Simply speaking, $L^*d\in R(L^*L)$ refers to existence of an LS solution, and existence of an LS solution implies existence of the pseudo-inverse. We will once again come back to this issue in Remark \ref{Na}.

Through the orthogonal projection to $N(L)^\perp, R(L)$ may be equipped with an inner product and so as to become a reproducing kernel Hilbert space (RKHS) with reproducing kernel $K(p,q)=\langle h_p,h_q\rangle_{{\mathcal H}}=K_p(q).$ To note this fact we  may write $R(L)=H_K.$ Note that here the space $H_K$ is induced from the operator $L,$ being identical in the set-theoretic sense with the range of $L,$ that is $R(L).$ Notation-wise, it is different from the space $H_K$ that we use in \S 1 as the domain of the operator $L.$ \\

Precisely, the inner product of $H_K$ is defined through
\[ \langle F,G\rangle_{H_K}\triangleq \langle P_{N(L)^\perp}f,P_{N(L)^\perp}g\rangle_{\mathcal H}, \quad Lf=F,\ Lg=G,\quad  \forall f, g\in {\mathcal H}. \]
The above construction gives rise to the relations
\[ Lh_p(q)=K_p(q), \ {\rm and}\ L: N(L)^\perp \to H_K\ {\rm is \ an \ isometric \ isomorphism}.\]
Due to this property $L$ restricted to $N(L)^\perp$ has a bounded inverse $L^{-1}: H_K\to N(L)^\perp.$

 Under the Hilbert-Schmidt condition $L$ is a bounded operator from ${\mathcal H}=L^2(D)$ to $H=L^2(\E).$ In fact, by first using the Minkowski inequality and then the Cauchy-Schwarz inequality, for any $f\in {\mathcal H},$
 \begin{eqnarray*}
  \|F\|_{L^2(\E)} &=&\left(\int_{\E}|\int_{D}f^+(x)\overline{h}_p(x)dx|^2dp\right)^{\frac{1}{2}}\\
  &\leq& \int_{D}|f^+(x)|\left(\int_{\E}|\overline{h}_p(x)|^2dx\right)^{\frac{1}{2}}dp\\
  &\leq& M\|f^+\|_{\mathcal H}\\
  &=& M\|F\|_{H_K}.
  \end{eqnarray*}
  This shows that the identical mapping from $H_K$ to $L^2(E)$ is an imbedding.

  We will frequently use the following properties in relation to a bounded linear operator in a Hilbert space and the adjoint operator in its conjugate Hilbert space.

  \begin{proposition}\label{prop}
  Let $L$ be a bounded linear operator from the Hilbert space $\mathcal H$ to the Hilbert space $H.$ Then
  \[ N(L)=R(L^\ast)^\perp \quad {\rm and}\quad \overline{R(L^\ast)}=N(L)^\perp=N(L^\ast L)^\perp=\overline{R(L^\ast L)}.\]
  \end{proposition}

  \noindent{Proof}. If $x\in N(L),$ then $L(x)=0.$ Hence for any $y\in H,$
  \[ \langle x, L^\ast y\rangle =\langle Lx, y\rangle=0.\]
  This implies $x\in R(L^\ast)^\perp$ and so $N(L)\subset R(L^\ast)^\perp.$ This argument is reversible, and we thus have $R(L^\ast)^\perp \subset N(L)$, and hence  $N(L)=R(L^\ast)^\perp.$ This implies $N(L)^\perp=(R(L^\ast)^\perp)^\perp=\overline{R(L^\ast)}.$\\
  It is easy to show $N(L)\subset N(L^\ast L).$ The reverse inclusion also holds as
  $L^\ast Lx=0$ implies $0=\langle x,L^\ast Lx\rangle =\langle Lx,Lx\rangle=\|Lx\|^2,$ and thus $Lx=0$ and $x\in N(L).$ Hence, $N(L^\ast L)=N(L).$ Since $L^\ast L$ is a bounded operator from $\mathcal H$ to $\mathcal H,$ we have $N(L^\ast L)^\perp=\overline{R(L^\ast L)}.$ The proof is complete.\\

Under such formulation we are to solve the problem (\ref{noRKHS}) with $H=L^2(\E).$ There are two cases: (i) $d$ is in $H_K;$ and
(ii) $d\in H\setminus H_K.$ In this section we are under the assumption that $R(L)$ is a closed subset of $H.$ In the matrix case this assumption is automatically fulfilled. \\

\noindent Case (i). In the case since ${\rm span}\{K_p\}_{p\in E}$ is dense in $H_K,$  for each $n$ we can find finitely many parameters $p_1^{(n)},\cdots,p_{k_n}^{(n)}$ and complex numbers $c_1^{(n)},\cdots,c_{k_n}^{(n)}$ such that in the $H_K$-norm,
\[ d=\lim_{n\to \infty}\sum_{j=1}^{{k_n}}c^{(n)}_jK_{p^{(n)}_j}.\]
Since the mapping induced by the mapping $K_{p^{(n)}_j}$ to $h_{p^{(n)}_j}$ is an isometric isomorphism between  $H_k$ and $N(L)^\perp,$ there exists $\tilde{f}$ such that
\[ \tilde{f}=\lim_{n\to \infty}\sum_{j=1}^{{k_n}}c^{(n)}_j h_{p^{(n)}_j}\]
and $L\tilde{f}=d.$ $\tilde{f}$ is of minimum norm since, in fact, $\{h_p\}_{p\in\E}\subset N(L)^\perp$ implies $\tilde{f}\in N(L)^\perp.$ \\

\noindent Case (ii). In this case the process is divided into two steps. The first is to obtain the direct sum decomposition $d=g+h, g\in \overline{R(L)}, h\in
\overline{R(L)}^\perp.$ Since ${R(L)}$ is closed, $\overline{R(L)}={R(L)},$ the problem is solvable as, in fact, the minimum norm least square solution is $L^{-1}(g) \in  N(L)^\perp$ and the least square mean distance from $d$ to $\overline{R(L)}$ is $\|h\|_{H}.$

Summarizing the above two case we conclude

\begin{theorem}\label{HHK}
In the ${\mathcal H}$-$H_K$ formulation if $R(L)$ is a closed subset of $H,$
then the LS problem (\ref{noRKHS}) always has a unique minimum norm LS solution, or a pseudo-inverse solution.
\end{theorem}

\begin{remark} If $R(L)$ is further a closed subspace of $H,$, then
the orthogonal projection $g$ into $R(L)$ may be obtained through
\[ \langle d,K_p\rangle_{H}=\langle g,K_p\rangle_{H}=\langle g,K_p\rangle_{H_K}=g(p).\]
In the case, with Matrix-POAFD algorithm, it is not necessary to compute out the projection $g.$ In fact, since $\langle d,K_p\rangle_{H}=\langle g,K_p\rangle_{H},$  one can just proceed the optimal parameter selections for $\langle d,K_p\rangle_{H}$ instead of $\langle g,K_p\rangle_{H}.$
\end{remark}

\begin{remark}\label{Na} The assumptions of closeness of $R(L)$ or $R(L)$ being a linear subspace of $H$ are just to ensure existence of pseudo-inverse for all $d\in H.$
In practice, neither closed-ness of $R(L),$ nor $R(L)$ being  a linear subspace of $H$ would be valid. In general situation the pseudo-inverse, or minimum norm least square   solution, is only available when $d$ has an incomplete direct sum decomposition \begin{eqnarray}\label{dagger}
d=g+h\in R(L)\oplus R(L)^\perp,\end{eqnarray}
where
$g\in R(L), h\in R(L)^\perp.$ With the incomplete direct sum decomposition the pseudo-inverse solution is $f=L^{-1}g$ and the LS value or the distance from $d$ to $R(L)$ is $\|h\|_H.$The following classical result asserts that existence of an $LS$ solution, existence of the pseudo-inverse, and existence of an incomplete direct sum decomposition of the objective function $b\in H$ are all the same thing.\end{remark}

\begin{proposition}\label{ad=Na}
The condition (\ref{ad}) and the condition (\ref{dagger}) are equivalent.
\end{proposition}

\noindent{Proof}. First assume (\ref{ad}) to hold. In the proof of Proposition \ref{LSexist} we showed in the case there exists $\tilde{f}$ such that $d=L\tilde{f} +(d-L\tilde{f})\in R(L)\oplus R(L)^\perp.$ On the other hand, $d\in R(L)\oplus R(L)^\perp$ implies $L^\ast d \subset R(L^\ast L)\cup L^\ast(R(L)^\perp)=R(L^\ast L).$
The proof is complete.\\

Either of the two equivalent conditions in Proposition \ref{ad=Na} is addressed as \emph{the pseudo-inverse criterion}. Two special cases are separately named:
The equation (\ref{noRKHS}) is called as a \emph{solvable equation} if $h=0;$ and a \emph{solvable projection equation } if $h\ne 0$.\\

\def\normal{{\rm normal}}
 Return to the matrix context. We formally define the set of the LS solutions and the set of the pseudo-inverse, the latter contains only a single element.
 \begin{definition}
 $$X^{-1}_{{\rm normal}}(\{Y\})=\{\tilde{W}\in {\R}^n\ :\ \|X\tilde{W}-Y\|=\inf_{W\in {\R}^n} \|XW-Y\|_{{\R}^m}\}$$
 and, if $X^{-1}_{{\rm normal}}(\{Y\})\ne \emptyset,$
 $$
 \{X^\dagger (Y)\}=\{\tilde{W}\ :\ \tilde{W}\in X^{-1}_{{\rm normal}}(\{Y\})\ {\rm and}\ \|\tilde W\|\leq \|W\|\ {\rm for\ any\ } W\in X^{-1}_{{\rm normal}}(\{Y\}).$$
 \end{definition}

Matric-POAFD, and other pseudo-inverse $X^\dagger (Y).$ As an application of ${\mathcal H}$-$H_K$ formulation we prove a result for computing the pseudo-inverse solution of (\ref{noRKHS}). We first consider  the matrix setting.
 The next theorem establishes
 the relation $\{X^\dagger (Y)\}=X^\ast{X^\ast}_{\normal}^{-1}\left(X_{\normal}^{-1}(\{Y\})\right).$ This relation shows that the pseudo-inverse can be obtained by solving consecutively two LS problems. In particular, by doing twice Matrix-POAFD, of which one is for the pair $(X,Y);$ and the other is for the pair $(X^\ast, W_1)$ for any $W_1\in X^{-1}_{\normal}(\{Y\})),$ one gets the pseudo-inverse for $(X,Y).$

\begin{theorem}\label{minimum norm 1}
 Let $W_1$ be any LS solution to $XW=Y,$ that is $XW_1=Y_1$ in the exact equality sense, where $Y_1$ is the projection of $Y$ into the span of the columns of $X.$ Let $W_2$ be any LS solution to
 \[ X^\ast W_2=W_1,\]
 then $X^\ast W_2$
 is the unique pseudo-inverse solution to $XW=Y.$ With the introduced notation there holds $\{X^\dagger (Y)\}=X^\ast {X^\ast}_{normal}^{-1}\left(X^{-1}_{\normal}(\{Y\})\right).$
 \end{theorem}

\noindent{Proof}. Under the ${\mathcal H}$-$H_K$ formulation it suffices to show that (i) the column vector $X^\ast W_2$ is an LS solution to $XW=Y;$ and (ii) $X^\ast W_2$ belongs to $N(L)^\perp,$ where the operator $L$ is the one defined by the matrix multiplication $LW=XW.$

 Firstly we show (i). From the pseudo identical relation $X^\ast W_2=W_1,$ we have the exact equality relation $XX^\ast W_2=XW_1,$ the latter gives rise to the exact relation $X(X^\ast W_2)=Y_1,$ where $Y_1$ is the projection of $Y$ into the span of the column vectors of $X.$ Hence, $X^\ast W_2$ is an LS solution to $XW=Y.$ (ii) reduces to showing that for any $V\in {\R}^n$ in the null space of $L,$ or equivalently, satisfying the relation $XV=0,$ there will hold $\langle X^\ast W_2,V\rangle=0.$ But this is obvious as $X^\ast W_2$ is a linear combination of the rows of $X,$ while $V$ is orthogonal to all rows of $X$ due to the relation $XV=0.$
The proof is complete.\\

In the matrix case the above two-step LS algorithm can be combined to form a one step LS algorithm.

\begin{corollary}\label{iter1}
The matrix pseudo-inverse solution for the equation $XW=Y$ is given by $X^\ast \tilde{W},$ where $\tilde{W}$ is any LS solution to the equation $\left(XX^\ast\right)W=Y.$
\end{corollary}

\noindent{Proof}. The LS problem for the matrix equation $\left(XX^\ast\right)W=Y$ is always solvable. The associative law of matrix product implies ${X^\ast}_{normal}^{-1}\left(X^{-1}_{normal}(\{Y\})\right)=(XX^\ast)^{-1}_{normal}(\{Y\}).$ The desired relation $X^\dagger Y=X^\ast \tilde{W}$ then follows from the proof of (ii) of Theorem \ref{minimum norm 1}. The proof is complete.\\

In the sequel we will call the algorithm corresponding to Theorem \ref{minimum norm 1} \emph{two-step Matrix-POAFD}, and that corresponding to Corollary \ref{iter1} \emph{one-step Matrix-POAFD}. Our tested examples, of which some are recorded in \S 5, show that the two-step and the one-step algorithms are particularly effective to the cases $m>>n$ and the $m<<n,$  respectively.\\

There exists a counterpart result for the functional space context. In the matrix case no assumptions are needed to be imposed for either the two-step algorithm (Theorem \ref{minimum norm 1} or the step algorithm for the product matrix $XX^\ast$ (Corollary \ref{iter1}). For the general Hilbert space case a condition stronger than $L^\ast d\in R({L^*L})$ needs to be imposed. First we give similar definitions for the LS and pseudo-inverse solutions with bounded linear operators in Hilbert spaces.

\begin{definition}
 $$L^{-1}_{{\rm normal}}(\{d\})=\{\tilde{f}\in {\R}^n\ :\ \|L\tilde{f}-d\|=\inf_{f\in {\mathcal H}} \|Lf-d\|_{H}\}$$
 and, if $L^{-1}_{{\rm normal}}(\{d\})\ne \emptyset,$
 $$
 \{L^\dagger (d)\}=\{\tilde{f}\ :\ \tilde{f}\in L^{-1}_{{\rm normal}}(\{d\})\ {\rm and}\ \|\tilde f\|\leq \|f\|\ {\rm for\ any\ } f\in L^{-1}_{{\rm normal}}(\{d\}).$$
 \end{definition}

\begin{theorem}\label{minimum norm 2}
For $d\in H$ satisfying $LL^\ast d\in R({LL^\ast LL^*})$ there holds
\begin{eqnarray}
L^\dagger(d)={L^\ast}{L^\ast}^{-1}_{\normal}L^{-1}_{\normal}(d)={L^\ast}(LL^\ast)^{-1}_\normal d.
\end{eqnarray}
\end{theorem}

\noindent{Proof}. Since $LL^\ast d\in R({LL^*LL^*}),$ we have, according to Proposition \ref{ad=Na},
\[ d=d_1+d_0\in R(LL^\ast)\cup R(LL^\ast)^\perp =R(LL^\ast)\cup R(L)^\perp\subset R(L)\cup R(L)^\perp,\] and, moreover,
 \[ d_1\in R(LL^\ast)\quad {\rm and}\quad d_0\in R(L)^\perp.\]
  This shows that $d$ satisfies the pseudo-inverse criterion for $Lf=d.$ In particular, there exists
$f_1$ such that, as equality relation, $L f_1=d_1.$ We next consider the second LS problem $L^\ast f=f_1.$ Due to the relations $Lf_1=d_1\in R(LL^\ast),$  the problem, too, has a solution (Proposition \ref{LSexist}), say $f_2,$ and therefore $L(L^\ast f_2)=Lf_1=d_1.$ This shows that $L^\ast f_2\in L^{-1}_{\normal}(d).$ To prove $L^\ast f_2$ is of minimum norm over all functions in $L^{-1}_{\normal}(d),$ in view of ${\mathcal H}$-$H_K$ theory, it suffices to show $L^\ast f_2\in N(L)^\perp.$ The last relation is equivalent to $\langle g, L^\ast f_2\rangle=0$ whenever $g\in N(L).$ This, however, is obvious due to the basic relation
\[ \langle g, L^\ast f_2\rangle = \langle Lg, f_2\rangle =0.\]
The proof is complete.

\begin{remark}
Theorem \ref{minimum norm 1} and Theorem \ref{minimum norm 2} propose an easy methodology in computing the Moore-Penrose generalized inverse: It is just by doing consecutively twice of any type of algorithm in finding least square solutions.  The formulas given in the theorems and the corollaries are not as  direct and explicit as that in (\ref{direct}). The latter is formulated with a RKHS and the reproducing kernel is indispensably involved.  Realization of an algorithm for formula (\ref{direct}) is by no means easy.  Crucially, in the process one has to work with an induced reproducing kernel $k$ of complicated form (see (\ref{k})). With the matrix case one can explicitly express the solution $f_d(p)$ by taking $K(p,q)=\vec{e}_p^\ast \vec{e}_q, p,q=1,\cdots,n,$ in formula (\ref{direct}), where $\vec{e}_p^\ast=(0,\cdots,0,1,0,\cdots,0),$ and $1$ appears in the $p$'s position. We omit the details that are not the main theme of the present paper.
\end{remark}

\section{Comparison Between Matrix-POAFD Algorithm and the Traditional Methods}

\subsection{Comparison of Efficiency of Solving LS Problem with Matrix-POAFD and Traditional Methods}
To compare efficiencies of different methods in solving the LS problem, we conducted a series of numerical experiments on synthetic datasets with varying matrix sizes and noise levels. Specifically, we randomly generated matrices $X,$ in which the columns are treated as features and the rows the individuals, and responding (observing) vectors $Y$, and compared four widely used feature selection methods: Least Absolute Shrinkage and Selection Operator (LASSO), Forward Selection (FS), Principal Component Regression (PCR), and Matrix-POAFD (two-Step). The experiments were designed to explore two key aspects: (i) The errors (Euclidean Norm) and computation times of the methods over matrices of varying dimensions. (ii)The robustness of each method under increasing noise levels. Over a number of experiments we recorded two cases: a $100 \times 10$ matrix and a $1000 \times 10$ matrix. Figures~\ref{f1} and~\ref{f2} respectively illustrate the comparison of errors and computation times across the four methods as the number of selected features increases for each case.

\begin{figure}[H]
    \centering
    \begin{minipage}{0.48\textwidth}
        \centering
        \includegraphics[width=\textwidth]{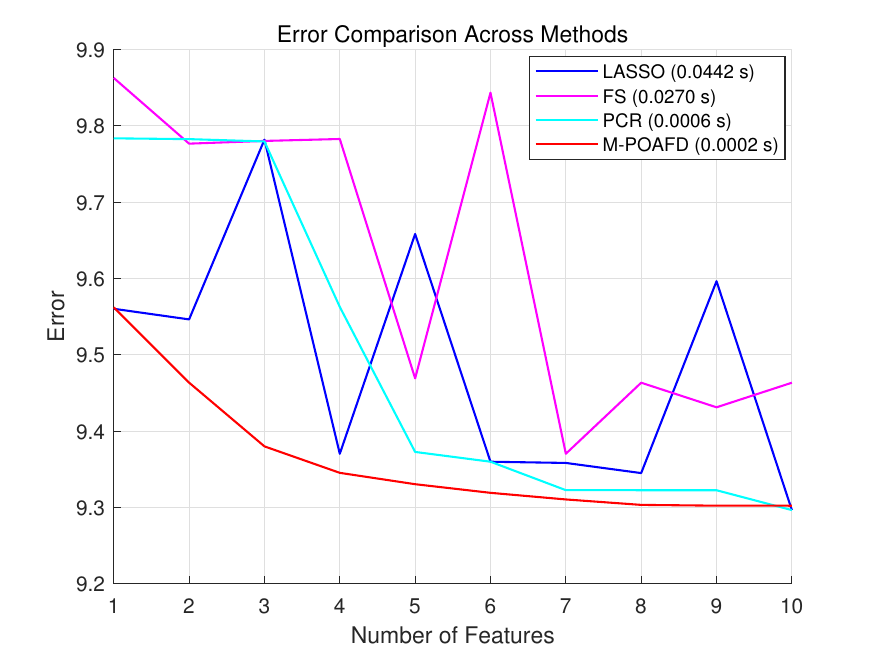}
    \end{minipage}
    \hfill
    \begin{minipage}{0.48\textwidth}
        \centering
        \includegraphics[width=\textwidth]{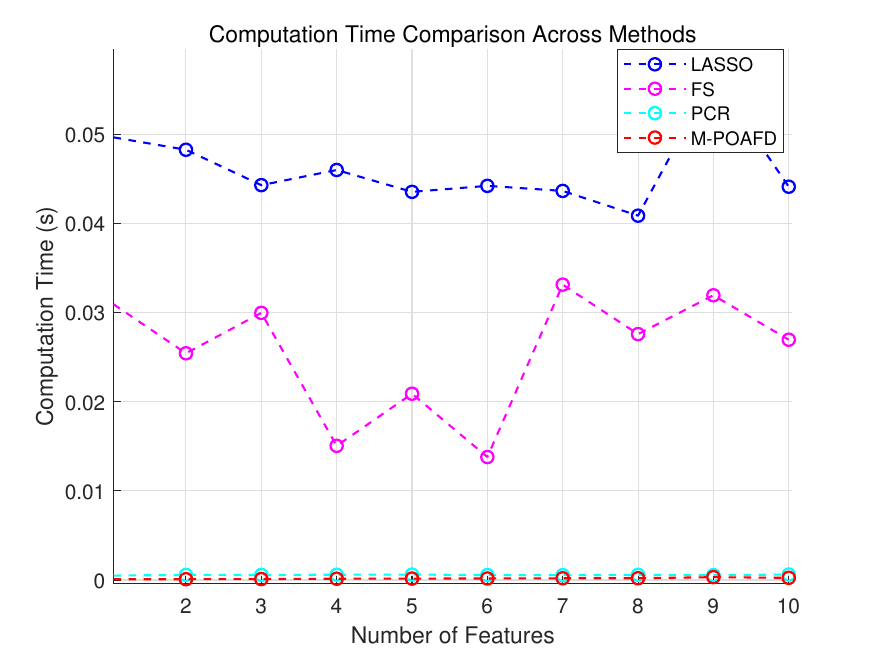}
    \end{minipage}
    \caption{Comparison of Error and Computation Time Across Four Methods on a \(100 \times 10\) Matrix with Increasing Number of Features}
    \label{f1}
\end{figure}

\begin{figure}[H]
    \centering
    \begin{minipage}{0.48\textwidth}
        \centering
        \includegraphics[width=\textwidth]{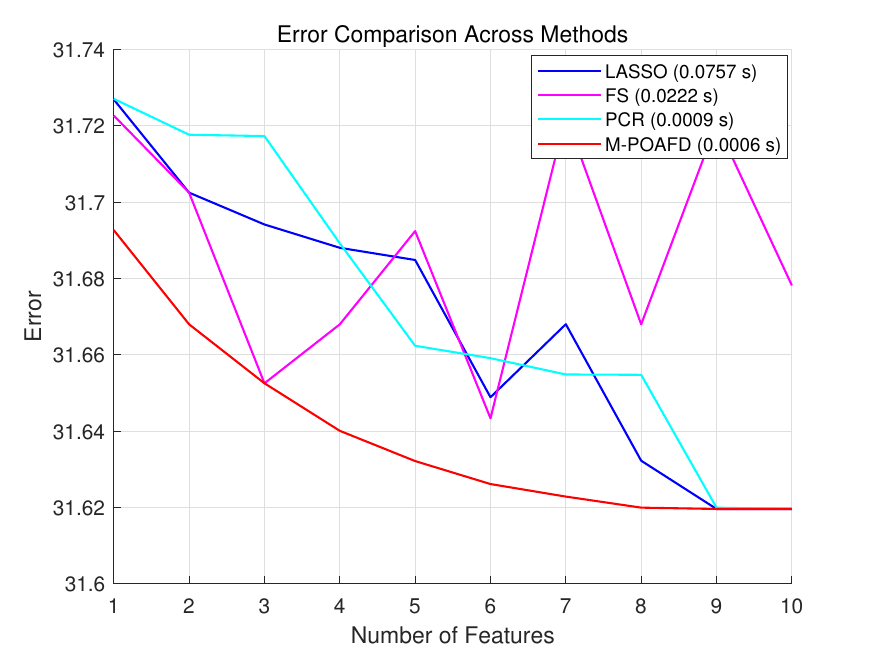}
    \end{minipage}
    \hfill
    \begin{minipage}{0.48\textwidth}
        \centering
        \includegraphics[width=\textwidth]{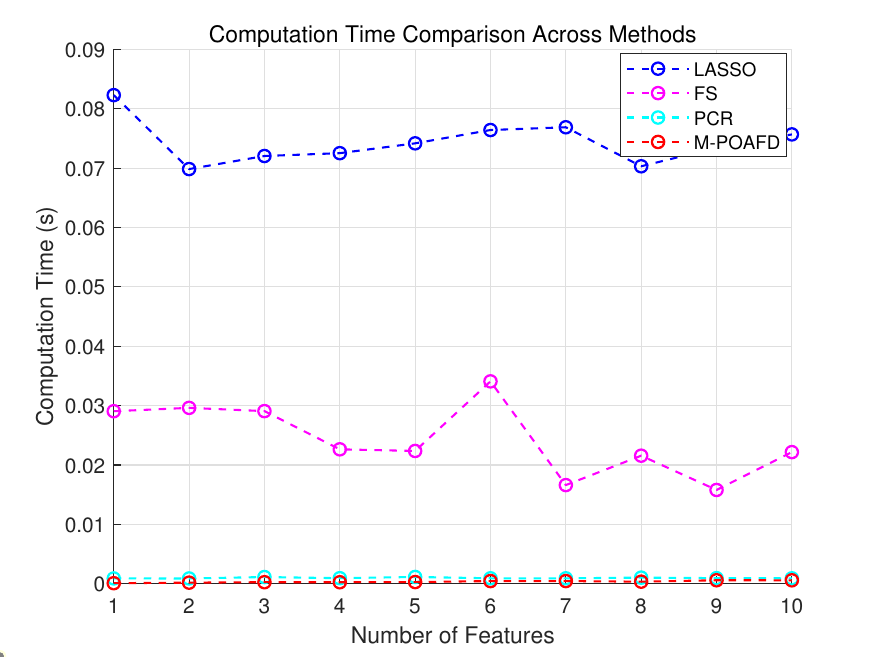}
    \end{minipage}
    \caption{Comparison of Errors and Computation Times Across Four Methods on a \(1000 \times 10\) Matrix with Increasing Number of Features}
    \label{f2}
\end{figure}
As shown in the left panels of Figures \ref{f1} and \ref{f2}, Matrix-POAFD stationally achieves lower reconstruction errors with regards to different matrix sizes, demonstrating its capability in extracting informative features. LASSO and PCR also exhibit reasonable performance, but the errors fluctuate remarkably noticeably. Forward Selection shows little reconstruction effect and tends to have higher variation in errors, likely being caused by the greedy nature of its feature selection process.

The right panels of Figures \ref{f1} and \ref{f2} show the comparison of computation times across different methods as the number of selected features increases. LASSO and FS exhibit moderate time variations due to the iterative nature of feature selection and the cross-validation steps involved in parameter tuning. In contrast, PCR and Matrix-POAFD (two-step) maintain relatively low and stable computation times, demonstrating their robustness and computational efficiency. Notably, Matrix-POAFD achieves the lowest and most consistent computation time, which can be attributed to its direct extraction of informative features based on the matching pursuit principle, avoiding the complexity introduced by cross-validation or greedy feature selection.
\\

To further examine robustness of different methods, we introduced Gaussian noise into the responding vector \( y \) of different levels. Specifically, we considered two cases: a moderate noise (\(\sigma = 0.5\)) and a higher level noise (\(\sigma = 5\)). Figure~\ref{f3} presents the error trends for each method under the tested noise levels.

\begin{figure}[H]
    \centering
    \begin{minipage}{0.49\textwidth}
        \centering
        \includegraphics[width=\textwidth]{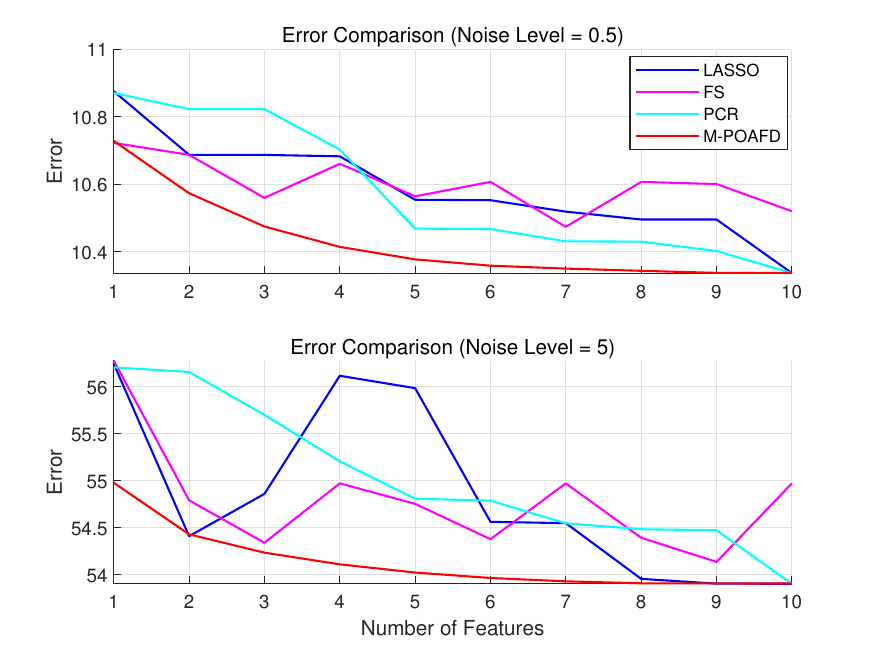}
        \caption*{\footnotesize\(100 \times 10\) Matrix}
    \end{minipage}
    \hfill
    \begin{minipage}{0.49\textwidth}
        \centering
        \includegraphics[width=\textwidth]{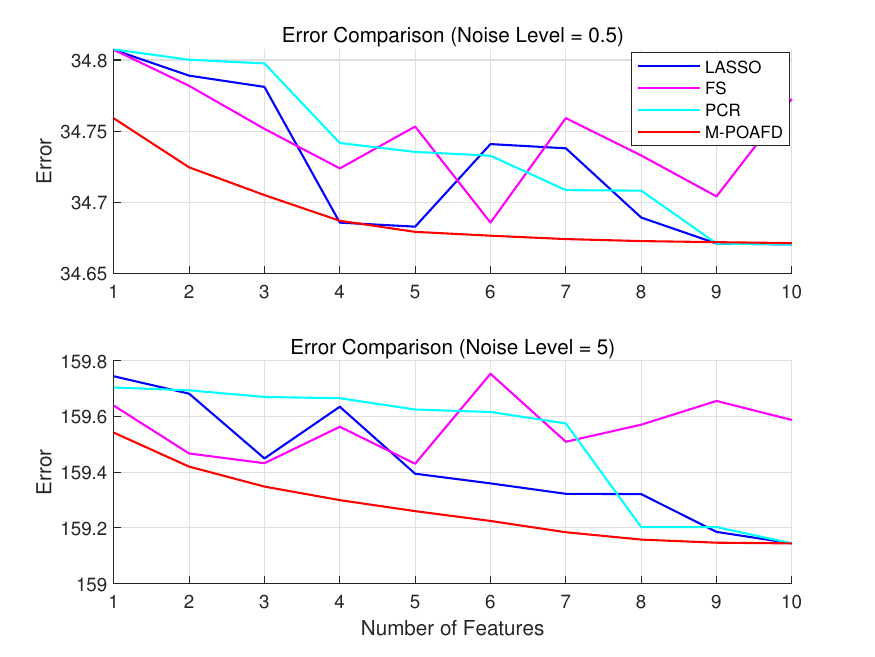}
        \caption*{\footnotesize\(1000 \times 10\) Matrix}
    \end{minipage}
    \caption{Error Comparison on \(100 \times 10\) and \(1000 \times 10\) Matrices Under Different Noise Levels}
    \label{f3}
\end{figure}

When the noise level is relatively low (\(\sigma = 0.5\)), all the methods show a general trend of decreasing error as more features are selected. With increased noise (\(\sigma = 5\)), the error magnitudes increase significantly for all methods. Nonetheless, Matrix-POAFD appears to be more stable, maintaining lower errors compared to other methods. LASSO and PCR show moderate performance, while Forward Selection exhibits the highest fluctuations, indicating its sensitivity to noise.

\subsection{Comparison of Efficiency of Solving Pseudo-Inverse Problem with Matrix-POAFD and Traditional Methods}

The LS solvers, including Matrix-POAFD, usually give solutions to LS problems. The obtained solutions are not necessarily of the minimum norm solutions, or, equivalently, not necessarily the pseudo-inverses of the problem. Theorem \ref{minimum norm 1} suggests a two-step algorithm that does give pseudo-inverse problems, in which each step can adopt any LS method. Corollary \ref{iter1} further establishes a one-step algorithm to solve pseudo-inverse solutions. The present section devotes to comparing several existing pseudo-inverse solvers with, in particular, the newly established two-step and the one-step Matrix-POAFD methods (2-M-POAFD, 1-M-POAFD), in terms of quantities of the minimum norms and the errors occurred. Six methods are compared in this study, including the Least Squares QR (LSQR), Conjugate Gradient (CG), Ridge Regression (Ridge), Moore-Penrose Pseudoinverse (MP), 2-M-POAFD and 1-M-POAFD. Figure~\ref{f4} presents comparison of the solution norms for two matrices: a tall \( 3000 \times 30 \) matrix (left) and a flat \( 30 \times 3000 \) matrix (right). In the tall matrix case, the two-step Matrix-POAFD algorithm yields the least minimum norm solution among the others by using almost the least computer running time, incurring comparable errors. Withe this $m>>n$ case the one-step Matrix-POAFD gives rise to excellent minimum norm LS solution with comparable errors, but with considerably longer computer running time. This last shortcoming is expected for it has to compute and do Marix-POAFD with the large matrix $XX^\ast$ of order $3000\times 3000.$ For the flat matrix, all methods yield almost identical minimum norms. Among which, however, the one-step Matrix-POAFD method requires the least time and achieves the highest accuracy. The miracle is due to the fact that the one-step method works with the small matrix $XX^\ast$ of order, merely, $30\times 30.$

\begin{figure}[H]
    \centering
    \begin{minipage}{0.49\textwidth}
        \centering
        \includegraphics[width=\textwidth]{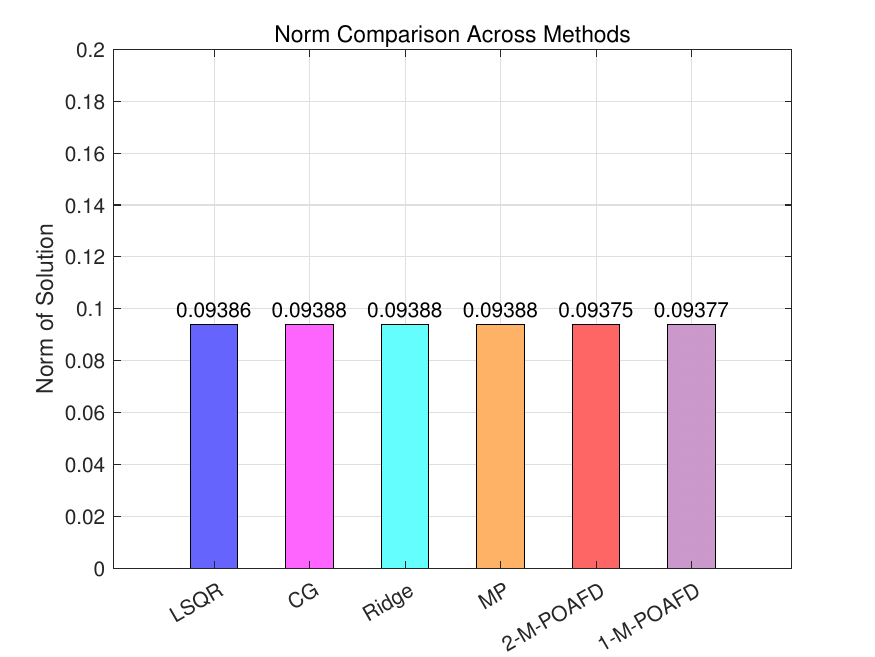}
        \caption*{\footnotesize\(3000 \times 30\) Matrix}
    \end{minipage}
    \hfill
    \begin{minipage}{0.49\textwidth}
        \centering
        \includegraphics[width=\textwidth]{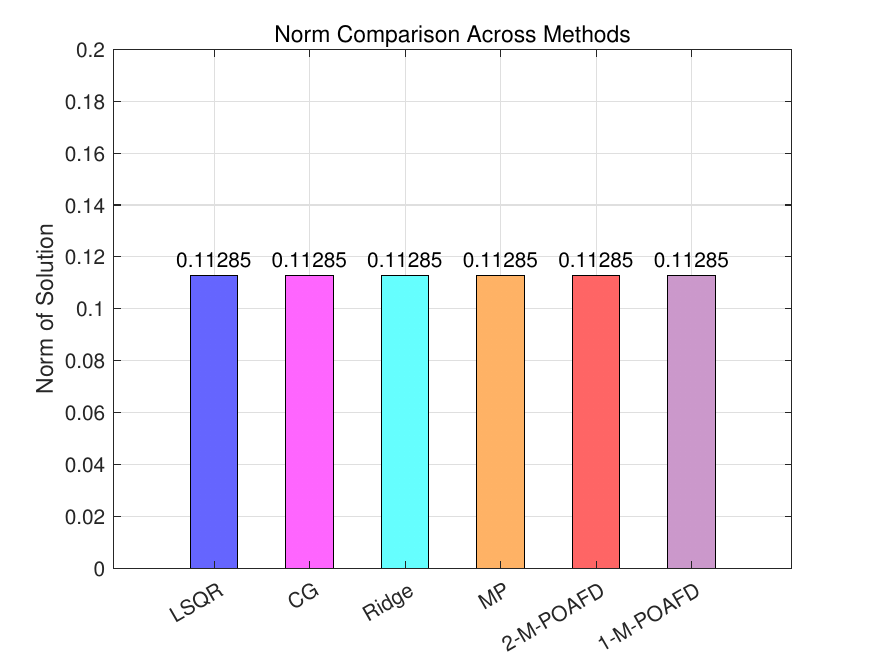}
        \caption*{\footnotesize\(30 \times 3000\) Matrix}
    \end{minipage}
    \caption{Comparison of Solution Norms Across Methods}
    \label{f4}
\end{figure}

In detail, the tables present a comparative analysis of computational time and reconstruction error across different methods for two measurement matrices: a tall \( 3000 \times 30 \) matrix (Table~\ref{t1}) and a flat \( 30 \times 3000 \) matrix (Table~\ref{t2}). The key observations are as follows. For the tall matrix, the CG method achieved the shortest solution time (0.0012 s), followed by the 2-M-POAFD method (0.0017 s). The 1-M-POAFD method exhibited the longest solution time (10.7254 s). Most methods (including LSQR, CG, Ridge, and MP) achieved similar errors around 54.2955, indicating consistency. On the wide matrix, the CG method's solution time increased significantly (0.0808 s), whereas the 1-M-POAFD method's solution time decreased drastically (0.0020 s). The 1-M-POAFD method demonstrated a remarkable numerical accuracy advantage, with an error as low as $1.8016 \times 10^{-15}$, highlighting its strong convergence properties for high-dimensional problems. The 2-M-POAFD method exhibited balanced performance between computational efficiency and error reduction across both types of matrix sizes. In contrast, the 1-M-POAFD method demonstrated exceptional accuracy for the flat matrix but required significantly more computational time for tall matrices.
In conclusion, the 2-M-POAFD method achieved a balanced trade-off between computational efficiency and error control, while the 1-M-POAFD method showed superior numerical stability and accuracy for, especially, $m<<n$ type problems.

\begin{table}[H]
    \centering
    \renewcommand{\arraystretch}{1.2} 
    \setlength{\tabcolsep}{10pt} 
    \begin{tabular}{ccc}
        \toprule
        \textbf{Method} & \textbf{Time (s)} & \textbf{Error} \\
        \midrule
        LSQR    & 0.0019  & 54.2955 \\
        CG      & 0.0012  & 54.2955 \\
        Ridge   & 0.0080  & 54.2955 \\
        MP      & 0.0024  & 54.2955 \\
        2-M-POAFD & 0.0017  & 54.2961\\
        1-M-POAFD & 10.7254  & 54.2961\\
        \bottomrule
    \end{tabular}
    \caption{\(3000 \times 30\) Matrix}
    \label{t1}
\end{table}

\begin{table}[H]
    \centering
    \renewcommand{\arraystretch}{1.2} 
    \setlength{\tabcolsep}{10pt} 
    \begin{tabular}{ccc}
        \toprule
        \textbf{Method} & \textbf{Time (s)} & \textbf{Error} \\
        \midrule
        LSQR    & 0.0020  & 0.0059 \\
        CG      & 0.0808  & 0.0059 \\
        Ridge   & 0.3036  & 0.0020 \\
        MP      & 0.0037  & 0.0015 \\
        2-M-POAFD & 0.0094  & 0.0008\\
        1-M-POAFD & 0.0020  & 1.8016$\times 10^{-15}$\\
        \bottomrule
    \end{tabular}
    \caption{\(30 \times 3000\) Matrix}
    \label{t2}
\end{table}


The numerical experiments conducted in this study provide a comprehensive comparison between Matrix-POAFD and most commonly used methods for solving least squares problems. Our results indicate that Matrix-POAFD consistently achieves solutions of smaller norms while maintaining competitive approximation accuracy. In the presence of noise, Matrix-POAFD demonstrates greater stability, yielding lower errors compared to other methods. Furthermore, our computational time analysis highlights that LSQR and CG are generally the fastest, whereas Ridge regression incurs higher computational costs. These findings would underscore the effectiveness of Matrix-POAFD in balancing solution stability, accuracy, and efficiency, making it a promising alternative for solving large-scale least squares problems.

\section*{Acknowledgement}
Sincere thanks are due to He ZHANG, for his helpful discussions and references during an earlier stage of this study.\par
 Chi Tin Hon and Tao Qian are supported by the Major Project of Guangzhou National Laboratory [grant number GZNL2024A01004] and the Science and Technology Development Fund of Macau SAR [grant number FDCT0128/2022/A, 0020/2023/RIB1, 0111/2023/AFJ, 005/2022/ALC]. Wei Qu is supported by the National Natural Science Foundation of China [grant number 1240010026] and the Zhejiang Provincial Natural Science Foundation of China [grant number LQ23A010014].

{}
\end{document}